\documentclass[twocolumn,superscriptaddress,amsfont,amssymb,amsmath, showpacs,balancelastpage]{revtex4-1} 
\usepackage{graphicx,longtable,mathrsfs,color,array}
\usepackage[hidelinks]{hyperref}
\usepackage[usenames,dvipsnames]{xcolor} 
\usepackage{amssymb,amsmath,mathtools,mathrsfs,slashed} 
\usepackage{epsfig,subfigure,placeins,float} 
\usepackage{booktabs,longtable,ctable,multirow} 
\usepackage{exscale,relsize} 
\usepackage[normalem]{ulem} 
\usepackage{enumerate}
\usepackage{times, mathptmx} 

\def\G{\Gamma}
\def\be{\begin{equation}}
\def\ee{\end{equation}}
\def\ba{\begin{eqnarray}}
\def\ea{\end{eqnarray}}

\def\L{\mathcal{L}}
\def\M{\mathcal{M}}
\def\R{\mathcal{R}}
\def\v{\mathfrak{v}}
\def\s{\mathfrak{s}}
\def\f{\frac}

\def\os{{}^{(0)}}
\def\on{{}^{(n)}}
\def\om{{}^{(m)}}
\def\p{\mathfrak{p}}
\begin{document}

\title{The Cosmology of an Infinite Dimensional Universe.}
\author{David Sloan }
\email{ David.Sloan@physics.ox.ac.uk}
\affiliation{Astrophysics, University of Oxford, DWB, Keble Road, Oxford OX1 3RH, UK}

\author{Pedro G. Ferreira}
\affiliation{Astrophysics, University of Oxford, DWB, Keble Road, Oxford OX1 3RH, UK}
\date{Received \today; published -- 00, 0000}

\begin{abstract}
We consider a universe with an arbitrary number of extra dimensions, $N$. We present a new method for constructing the cosmological equations of motion and find analytic solutions with an explicit dependence on $N$. When we take the $N\rightarrow\infty$ limit we find novel, emergent behaviour which distinguishes it from normal Kaluza-Klein universes. \end{abstract}


\maketitle

\section{Introduction}
\label{sec:intro}

 The observational evidence for life in 3D is unassailable. There have been attempts at justifying why this is so, i.e. why a three dimensional universe is natural or inevitable. Ehrenfest famously argued that  it was impossible to construct stable classical orbits in higher dimensional spaces \cite{ehrenfest1917way}. Conversely, in less than 3 dimensions, there is no gravitational force (or, to be correct, it is topological in nature) \cite{Deser:1983tn}.  
 
Our three dimensional space could be a slice of a higher dimensional space-time \cite{Deser:1998ed}. What the exact number of extra dimensions is, remains to be determined. String theory has a preference for 11 or 26 dimensions. Indeed it has been proposed that it is through the interaction of the extra dimensions and the statistical mechanical properties of fundamental strings that three large dimensions naturally emerge \cite{Brandenberger:1988aj}.  Some of the more exotic theories of the multiverse allow for an arbitrary number of extra dimensions breathing in and out of existence. 

It is fair to say that a certain agnosticism prevails with regards to the number of extra dimensions although pragmatism tends to favour less rather than more \cite{Tegmark:1997jg}. The classic model of Kaluza and Klein posits the existence of one extra dimension and attempts to tie it to the vector potential of the electromagnetic source. The more recent brane-world universes \cite{ArkaniHamed:1998rs} tend to reside in 4+1 dimensional spacetimes with various schemes for deforming, or
"warping" the extra dimension \cite{Randall:1999ee} . 

Very little attention has been paid to case in which one has a large number of extra-dimensions, $N$. The expectation is that such a scenario might be calculationally unwieldy or that the dynamics of the extra-dimensions will completely overwhelm the dynamics of the full space-time. What little work there has been has focused on small scales. Strominger pioneered efforts by attemping to calculating scattering processes using an expansion in $1/N$, very much along the lines of what has been done for $SU(N)$ \cite{Strominger:1981jg}. The mantle has been picked up  by a few authors, focusing on the field theoretic and quantum properties of Kaluza-Klein universes \cite{Canfora:2005ax}, asymptotic safety\cite{Litim:2003vp,Fischer:2006fz}, non-local gravity\cite{Conroy:2015nva}, quantum gravity in general \cite{BjerrumBohr:2003zd} and on a lattice \cite{Hamber:2005vc}. For a historical perspective, see \cite{Deser:2004sz}.
The study of the classical properties arising from many extra dimensions has focused on black holes. In \cite{Emparan:2013moa}, the authors explored the fact that for large $N$, the gravitational force becomes ultra-localized. They have used this property to explore the fact that  black holes become non-interacting and have looked at how results in the gravitational canon (such as instabilities, wave equations on curved background, radiation, etc) are modified in the $N\rightarrow \infty$ limit. 

While the focus of previous work on $N\rightarrow \infty$ has been on small scales, we wish to look large and explore the consequences for cosmology. To do so, we will study homogeneous, but not necessarily isotropic, universes in the case of large $N$ with an eye on finding meaningful solutions in the infinite dimensional case.

We structure the paper as follows. In Section \ref{sec:eom} we present a novel technique for constructing the field equations in arbitrary dimensions and determine the equations of motion. In Section \ref{sec:solutions} we find solutions for arbitrary $D$ and $N$ for the vacuum, isotropic matter (including the cosmological constant) and anisotropic matter. In Section \ref{sec:infinity} we explore the $N\rightarrow\infty$ limit of our solutions and discuss their phenomenology. In Section \ref{sec:discussion} we conclude.
 
Throughout this paper we work with $8\pi G=1$ and the mostly plus metric signature.  The time direction is the zero coordinate, lower case Roman letter run over the background space, greek letters over the extra dimensions, and upper case Roman letters over all spatial dimensions. 
\section{Setup}
\label{sec:setup}

We will consider space-time manifolds of the form $I \times \M_1 \times \M_2 \times ... \times \M_q$ in which $I$ is an interval of $\mathbf{R}$ which comprises our time direction, and $\M_i$ are homogeneous spatial manifolds of dimension $D_i$. This splitting is motivated by the idea that the physically relevant cosmology is that of a large three dimensional spatial manifold and a (potentially large) number of small extra dimensions. With this in mind, we consider metrics of the form:
\be ds^2 = -dt^2 + e^{\frac{2v_1}{D_1}} ds_1^2 +e^{\frac{2v_2}{D_2}} ds_2^2 + ... + e^{\frac{2v_q}{D_q}} ds_q^2  \ee
Wherein $v_i$ are functions of time representing the logarithms of the volumes of the homogeneous manifolds, and the $ds_i$ are the line elements of the unit homogeneous manifolds. The usual technology of Kaluza-Klein compactifications is to begin with a large manifold of arbitrary dimension and perform a compactification of a number of these to leave a manifold with a mixture of `large' and 'small' dimensions. Our motivation is somewhat different - since we want to be able to take a well-defined limit in which the number of dimensions is taken to be infinite, we do not want to begin with an infinite dimensional manifold and follow this process. Rather we consider the effect of taking the product of a space-time with a (series of) homogeneous manifold(s) which have small volume and whose dimension is taken to infinity. Further our choice of considering the metric written in terms of the volumes of homogeneous manifolds (rather than their scale factors) may seem unusual; one normally considers a `compactification length' when making reductions. However, as we shall see below, when taking the limit as the number of small dimensions tends to infinity any non-infinitesimal change in the scale factor of an infinite dimensional manifold will completely dominate dynamics to the point that all the remainder of the dynamics is rendered trivial. 

For metrics of this form we find that the Ricci scalar, $R$ splits into components ${}^n R$, being the Ricci scalars of $\M_n$, a kinetic component from each of the volumes, and a cross-term ;
\be R = \sum_{i=1}^{q} \left[e^{-\frac{2 v_i}{D_i}}  ({}^i R) + 2 \ddot{v_i}+ \frac{D_i+1}{D_i} \dot{v_i}^2 \right]  + 2 \sum_{a \neq b} \dot{v_a}\dot{v_b} \ee
We note here that ${}^i R$ are simply numbers dependent on the topology of the unit volume manifolds $\M_i$ - for example if $\M_j=S^l$, the $l$-dimensional sphere, then ${}^j R = l(l-1)$. As we are dealing with the product of homogeneous manifolds, there is only one independent diagonal element of the rank $(1,1)$ Einstein tensor on each manifold. We shall label the spatial eigenvalue of the Einstein tensor of the space-time $I \times \M_n$ by ${}^n G$. For the element whose corresponding eigenspace covers $\M_i$ we find;
\be G_i = {}^i G + \sum_{p \neq i} \frac{e^{-\frac{2v_p}{D_p}}  ({}^p R)}{2} - \frac{\ddot{v_p}}{2} - \frac{D_p+1}{2D_p} \dot{v_p}^2 - \sum_{a < b} \dot{v_a}\dot{v_b} -  \frac{\dot{v_i}}{D_i} \sum_{p \neq i} \dot{v_p} \label{Pressures} \ee
Note here that Einstein's equations read $G_n = P_n$ where $P_n$ is the pressure arising from the stress-energy tensor. The final eigenvalue of the Einstein tensor, the energy density $G_o = \rho$ is the Friedmann equation;
\be \rho = \sum \rho_i + \sum_{a < b} \dot{v_a}\dot{v_b} \label{Friedmann} \ee
wherein $\rho_i$ is the energy density of the spacetime $I \times \M_i$. 


\subsection{Lagrangian Description}

The gravitational Lagrangian is $\L_g = \sqrt{g}R$. The splitting that we have performed above allows us to express this as the Lagrangian for one of the spatial manifolds with extra terms arising from the others; this is particularly useful when considering the effects of small extra 'internal' dimensions on the evolution of a large background geometry. Here we will single out $v$ as being the volume of the large background space, and label the logs of internal manifold volumes as $u_i$ of dimension $N_i$, of total log volume $u=\sum_i u_i$. To make the comparison with GR in $3+1$ dimensions clear, we will express the background space-time Ricci scalar as $\R$.
\be \L_g = e^{v+u} \left[\R + 2\dot{v}\dot{u} + 2\ddot{u} +\sum_i  \left( \frac{N_i+1}{N_i} \dot{u_i}^2 + e^{-\frac{2u_i}{N_i}} ({}^iR) \right) \right] \label{UVLag} \ee
Thus we see that the internal dimensions can be cast in the role of a non-minimally coupled scalar field - by making the transformation $\phi_i = \log(u_i)$ we could rewrite this Lagrangian in the familiar Brans-Dicke form with a non-standard kinetic term, and a coupling between the kinetic term and the Hubble rate of the large dimensions. It is further apparent that should the dynamics stabilize the extra dimensions ($\dot{u}=\ddot{u}=0$) then we will recover the Einstein-Hilbert Lagrangian for the background manifold, albeit with a modified coupling to matter, and the curvature term from the extra dimensions appearing as an effective cosmological constant.

\section{Infinite Dimensional Limits}
\label{sec:infinite}

Before continuing to show equations of motion and their solution in some physically interesting cases, it is important to note certain choices in our setup which affect the manner in which we will take the $D_i \rightarrow \infty$ limit. In particular, we have chosen to explore the dynamics of our system in terms of the logarithmic volumes of the spatial manifolds, as we are interested in the case in which the dimension of one of these manifolds is taken to infinity. There is a simple connection between these and the logarithmic scale factors; $v_i = D_i a_i$. However, when we take the infinite dimensional limit, we note that the expression in terms of volume captures a more subtle set of dynamics than that of scale factor alone, since $\delta v_i = D_i \delta a_i$ dictates that any non-zero change in scale factor be reflected by an infinite change in volume, and similarly any finite change in volume is infinitesimal in scale factor. For example the case where $D_k \rightarrow \infty$ any finite change in the corresponding scale factor $a_k$ would force the rest of the $a_i$ to constants in order to satisfy the Friedmann equation \ref{Friedmann} with finite energy density. Similarly, this is manifest in the gravitational Lagrangian \ref{UVLag}; each of the kinetic terms comes with a pre-factor that ranges from $1/2$ (in the case of two dimensions) to $1$ in the infinite limit.  Had we instead considered scale factors, these factors would have contained a term proportional to $D_i^2$, and hence in any of the equations of motion, this term would completely dominate the dynamics, fixing the scale factor for $\M_i$ and giving no further dynamical information. 

Another way in which one could take the infinite dimensional limit is to take the limit where the number of spatial manifolds becomes large. This is a physically distinct scenario from taking one of the manifolds large, as the behaviour of the Ricci scalar of the manifolds contributes terms that depend on the dimension of the manifold. This is particularly clear when we consider, for example $S^D$, as a spatial manifold; $R(S^D)=D(D-1)$, and hence grows to dominate dynamics completely in the infinite dimensional limit. Note that if our internal manifold is simply connected, the Killing-Hopf theorem states that it must be the $D$-sphere, $S^D$, $D$-dimensional euclidean space $E^D$ ($R=0$), or the $D$-dimensional hyperbolic plane, $H^D$ ($R=-D(D-1)$). In the non-flat cases, therefore the curvature dominates dynamics entirely and the system becomes frozen. We could further consider the case in which the internal manifold is formed as the infinite product of lower dimensional manifolds. The Ricci scalar of a product manifold is the sum of Ricci scalars (i.e. $R(\M_1 \times \M_2) = R(\M_1) + R(\M_2)$). To explore this situaiton, consider for example the contribution of $q$ identical manifolds each $S^n$ for some $n$ which we describe in terms of the volume of each spatial manifold, $v$. Then the contribution of this to equation \ref{UVLag} becomes:
\be \L_g = e^{v+ q u} \left[\R + 2q\dot{v}\dot{u} + \frac{q(qn-1)}{n} \dot{u}^2 + q e^{-\frac{2u}{n}} n(n-1) \right] \ee
Thus the dynamics of the internal manifold will again dominate that of the background. It would further appear that the kinetic term overwhelms the potential (curvature) term in the $q \rightarrow \infty$ limit, and the system would be independent of the choice of topologies. This highlights the subtlety involved in the choice of variables - had we instead chosen to express the dynamics in terms of the total volume of this manifold, $V=qv$ we would see that the potential term dominates the kinetic. In either case, the dynamics of the background manifold is rendered trivial as a result of the infinite dimensional product. Since the Ricci scalar is the product of sectional curvatures and our spatial manifolds are homogeneous and isotropic, any curvature of the internal infinite dimensional manifold is incompatible with a dynamical background manifold. One could consider the case in which our internal manifold is anisotropic, and thus can consist of the product of a finite dimensional curved space and an infinite dimensional flat space (e.g. $S^D \times E^q$ with the limit $q \rightarrow \infty$ taken). However in order to provide tractable equations, for the remainder of this article we will restrict ourselves to the product of flat manifolds.

\subsection{Flat Spatial Slices}

When considering the spatial manifold to consist of a product of tori, our system simplifies since for these spaces the Ricci scalar is dependent only on the Hubble rates, not the physical volumes. This is unsurprising as the geometry of flat spaces picks out no preferred length scale, as there is no radius of curvature. Although one might consider that the length scale around a torus would be a natural scale, the condition of homogeneity already identifies all points on the manifold, and therefore no direct physical role for the volume (or scale factor) to play. For a torus the Ricci scalar of the spatial manifold is zero and thus we find the dynamics of a space-time wherein the spatial manifold consists of the product of $q$ tori of dimensions $D_i$ is governed by a Lagrangian density consisting only of kinetic terms. This is given (up to a total divergence term introduced to render the system first order)
\be \mathcal{L}_g =  e^{\sum_i v_i} \left( \sum_i  \frac{D_i-1}{D_i} \dot{v_i}^2+ 2\sum_{i \neq j} \dot{v_i} \dot{v_j} \right) \ee
Note that in this action the only dependence on the volumes $v_i$ of the tori is in an overall factor. We could absorb this into the lapse to work in conformal time rather than proper time, however for clarity of exposition we will not do so. The gravitational Lagrangian can be minimally coupled to a matter Lagrangian which is also independent of the anisotropies (shape parameters) $\s_i$ for, e.g. perfect fluids on such a space-time, with the Hamiltonian giving rise to the Friedmann equation (\ref{Friedmann}). As such the dynamics of the system break into those of the total volume $\v=\sum_i v_i$ and the behaviours of the anisotropic expansions which determine the evolution of the shape of the tori, $\s_i$. Our Lagrangian only contains time derivative terms in $\s_i$ hence we find that there will be $q-1$ conserved quantities. 

This splitting into shape and volume terms is the basis of the ``cosmological billiards'' approach to homogeneous vacuum GR \cite{Heinzle:2007kv,Damour:2002et}. In such approaches the dynamical system is described in terms of the log volume and ``Misner parameters'' which describe the relative anisotropies. The Ricci gives rise to a potential which can be split as the product of that for the volume and the shape potential, which asymptotes to, e.g. a sharp-walled triangular well in the 3D case. Recent work in the relational description of such a system has shown that it retains a well defined dynamics of shape through the singularity even though the volume becomes singular \cite{Koslowski:2016hds}. The cosmological billiards paradigm has proven particularly useful in examining the dynamics of low dimensional curved space-times close to singularities (particularly in light of the BKL conjecture). However, in the case of flat spatial slices the shape potential is exactly zero, and thus the billiard dynamics is always that of simple linear motion. For completeness, we develop the generalized billiards approach to our space-times in appendix \ref{billiards}.

These equations reproduce those in \cite{Alvarez:1983kt}, for example, if we assume that the we have the product of two flat Robertson-Walker metrics. However our expression is much more general - we can consider the product of any number homogeneous manifolds.

\subsection{Matter}

As our cosmology is to be homogeneous and isotropic on submanifolds, we will consider matter that is subject to the same restrictions. In such a case, the high dimensional limits of many types of matter will be identical. As an example, consider radiation whose energy density is reduced by both dilution of the fields and redshift; in the isotropized limit of an $N$-dimensional spatial slice, the energy density will follow $v^{-(1+\frac{1}{N})}$ and thus in the high $N$ limit, will be indistinguishable from dust. A more accurate model of radiation would consider the matter as an anisotropic fluid as only expansion in the direction of the radiation's velocity will be redshifted; at small scales this plays a significant role \cite{Sloan:2016efd}. A similar argument applies for the isotropized limits of most matter types; in the infinite dimensional limit, the dilution factor is all that remains, thus for isotropized matter sources, $w \rightarrow 0$ as $N\rightarrow \infty$. The exceptions to this are the stiff fluid/massless scalar field, for which $\rho \sim v^{-2}$ and hence $w=1$, and the cosmological constant, $\rho \sim \Lambda$, $w=-1$. Hence in the isotropic limit, these are the only interesting matter cases; the $1/N$ contributions from redshift etc to the dust behaviour do not play a role in the limit. 

When considering anisotropic matter, the system is more complex. Particularly, we consider matter that behaves differently on the internal manifolds from the external. An apparent example is that of a $\p$-brane which wraps around $\p$ of the compact internal dimensions. As such, the winding modes of the brane contribute pressures given $w_\p = -\frac{\p}{D}$ in the directions in which they wrap, and $w_x$ determined by the fields on the brane in the remaining directions \cite{Brandenberger:2001kj}. We can therefore reconstruct matter for which the averaged anisotropic pressures take any values in $[-1,1]$. 

\section{Equations of Motion} 
\label{sec:eom}

In the special case of two tori of dimension $D$ and $N$ respectively, with logartihmic volumes $v$ and $u$, the Einstein equations are
\ba  \ddot{v}+\f{D+1}{2D} \dot{v}^2 + \f{N-1}{N} \dot{v}\dot{u} +\f{N-1}{2N} \dot{u}^2 + \f{N-1}{N} \ddot{u} &=& -P_x \label{eq:eom2} \\  
       \f{D-1}{D} \ddot{v}+ \f{D-1}{2D}\dot{v}^2 + \f{D-1}{D} \dot{v}\dot{u} +\f{N+1}{2N} \dot{u}^2 + \ddot{u} &=& -P_y \label{eq:eom1} \ea
with Hubble equation given by
\be \f{D-1}{2D} \dot{v}^2 + \dot{v}\dot{u} + \f{N-1}{2N} \dot{u}^2=\rho \label{eq:en1}\ee
We can see in equation \ref{eq:en1} that for large $N$, $\dot{v}$ and $\dot{u}$ must either have the same sign or there must be a large and consistent hierarchy in expansion rates  for the energy density to remain positive. Note too that the topology of our system ensures that there are no raw $v$ or $u$ terms in the field equations only in the $P_\alpha$ and $\rho$ terms. Furthermore, equations \ref{eq:eom1} are identical under swapping the $N$ for $D$ and $v$ for $u$ simultaneously, as expected - at this point we have no preference between the two. 

If we now consider isotropic pressures, $P_x=P_y$,  for example, and difference the two equations of motion, we are left with 
\be \f{1}{D} \ddot{v}+ \f{1}{D} \dot{v}^2 + \f{N-D}{DN} \dot{v}\dot{u} - \f{1}{N} \dot{u}^2 - \f{1}{N} \ddot{u} = 0 \ee
which can be written as a total derivative
\be e^{-(v+u)} \f{d}{dt} \left[e^{v+u} \left(\f{\dot{v}}{D}-\f{\dot{u}}{N} \right) \right] = 0 \label{eq:td}\ee
One immediate implication is that isotropic matter will lead to isotropic expansions at large volumes; when $u+v$ is large, the directional Hubble rates become equal. If we change the time variable to $d\eta = e^{-(u+v)} dt$, with derivatives by $\eta$ denoted by a prime, we end up with a constant of motion
\be \f{v'}{D}-\f{u'}{N} = C \label{eq:const}\ee
Note that we can rescale $\eta$ to set $C=1/D$ which simplifies calculations further down the line. We do not consider the cases where $C=0$ as these would constitute perfectly isotropic space-times, and thus are simply the $(N+D)$-dimensional extensions of the Friedmann-Lem\^aitre-Robertson-Walker cosmologies. The relationship between $t$ and $\eta$ introduced by changing our time variable is monotonic since the exponential function is everywhere positive for real arguments. We therefore cover the entire history of the universe in proper time under this transformation. In other words, $\eta$ is a good clock for our system. 

These results generalize to a spatial manifold consisting of the product of any number of tori with (possibly) different volumes. For isotropic matter the pressures are equal, and thus all the eigenvalues of the Einstein tensor are equal. Thus from equation \ref{Pressures} we find
\be e^{-\sum v_k} \frac{d}{dt} \left[e^{\sum v_k} \left(\frac{\dot{v_i}}{D_i} - \frac{\dot{v_j}}{D_j} \right)\right] = 0 \ee
We follow the analogue of the above change of time,  $d\eta = e^{-\sum v_i } dt$ and thus arrive at a set of conserved quantities;
\be \frac{v_i'}{D_i} - \frac{v_j'}{D_j} = C_{ij} \label{Constants} \ee
Note that the $C_{ij}$ comprise a space of $q-1$ independent components, as $C_{ij} = C_{[ij]} $ and $C_{ij}=C_{ik} + C_{kj}$. 

We can now use  equations \ref{eq:td} and \ref{eq:const} to explore what happens in some simplified scenarios.

\section{General Solutions.}
\label{sec:solutions}

Let us first try and find the vacuum solution. Note that the perfectly isotropic vacuum in any dimension is Minkowski space. To see see this, one has that, from isotropy,  the integration constant $C$ is exactly zero, forcing $u'=0$ for any non-zero $v'$. But equation \ref{eq:en1} has $\rho=0$ which implies $v'=0$  hence we have the isotropic Minkowski vacuum. 

A more interesting result arises in the case of the anisotropic vacuum which is analogous to the Bianchi I type solution with one degree of anisotropy in 3-dimensions, taking our spatial manifold to be the product of two tori. To solve the equations we first note that the conserved quantity in equation \ref{eq:const} can now be combined with equation \ref{eq:en1} to give us a quadratic equation for either $u'$ or $v'$ and will give us possible solutions for the vacuum. Setting $C=1$ for convenience we have $v'=1+\f{D u'}{N}$ and 
\be u'= \f{-N \pm \f{N^{3/2}}{\sqrt{D(D+N-1)}}}{D+N} \ee

We find then that the vacuum solutions (i.e. the Kasner solutions) always give rise to power law scale factors - hence our spacetimes take the form:
\be ds^2 = -dt^2 + t^{n_x} d\vec{x}^2 + t^{n_y} d\vec{y}^2 \ee
with the two solutions for each pair of $n_x$ and $n_y$ given by the choice of sign in the square root. These two powers are:
\ba n_x&=& \f{1 \pm \sqrt{\f{N}{D}(N+D-1)}}{D+N} \nonumber \\ n_y&=& \f{1\mp \sqrt{\f{D}{N}(N+D-1)}}{D+N} \ea

These correspond to the specific Bianchi I LRS cases of higher dimensional cosmology noted in other contexts\cite{Barrow337,Feinstein:1999ij} . One can easily verify that these two solutions in the $D=1$, $N=2$ case correspond to the two locally rotationally symmetric Bianchi I vacuum solutions with exponents either $\{1,0\}$ or $\{-\f{1}{3},\f{2}{3}\}$. Readers familiar with Bianchi I solutions will note that the anisotropic shear follows the $t^{-2}$, and the total spatial volume $t$ as we would expect. Hence the effective Friedmann equation for the isotropic (geometrically averaged) scale factor $\bar{a}$ is 

\be \f{\dot{\bar{a}}^2}{\bar{a}^2} = \frac{\Sigma}{\bar{a}^{2(D+N)}} \label{eq:avsf}\ee

To find the vacuum solutions with an arbitrary number of anisotropies, it is sufficient to find the solution in the case where the spatial manifold consists of $q$ anisotropic copies of $S^1$, and impose equalities on some of the volumes after finding the solution. This simplifies the algebra considerably, as $D_i=1$ throughout. The Friedmann equation \ref{eq:en1} reads simply 
\be \sum_{i < j} \dot{v_i} \dot{v_j} =0 \ee
which can be solved together with the Einstein equations to again give power laws for the scale factors - with metric
\be ds^2 = -dt^2 + t^{2P_i} dx_i^2 \ee
wherein the $P_i$ are constrained $\sum P_i = \sum P_i^2 =1$.


Let us now turn to isotropic polytropic, matter, with $P_x=P_y\equiv P=w\rho$. As our primary interest will be in the $D_j \rightarrow \infty$ limit for some choice of $j$, this will inform our strategy for examining the equations of motion; from \ref{Constants} it can be seen that if only one of the internal manifolds is to have its dimension taken to be large, the remaining spatial manifolds (including the large external manifold) will have their volumes evolve trivially in $\eta$ - described by $v_i'=C_i$ for some constants $C_i$. Let us first examine the case in which there is only one internal manifold. We will label its volume $u$ and dimension $N$; we will first solve the dynamics of the $u$ as our conserved quantity (equation \ref{eq:const}) renders $v$ trivial in this limit. Defining $\xi=w-1$ we can use
equation \ref{eq:const} to replace $v'$ and find 
\be u'' + \left(\f{N+D}{2N} \right) \xi u'^2 \nonumber + \xi u' +  \left[\f{N(D-1)}{2D(N+D-1)}\right] \xi  = 0 \label{eq:iso1}  \ee
We note from this that, if there exists a turning point in $u$, given that $\xi \leq 0$, it will be a minimum. Therefore we see that outside the case of the stiff fluid ($\xi=0$) we cannot stabilize the volume of the extra dimensions with an isotropic fluid. Furthermore, as a mixture of fluids can be given an effective equation of state no mixture of isotropic fluids can achieve stability of the extra dimensions. 
For the case in which $\xi=0$ - the stiff fluid, the equations of motion become trivial, and we 
quickly recover the solution $u=A \log(t)$, $v=(1-A)\log(t)$. Thus we see that in the isotropized effective Friedmann equation \ref{eq:avsf}, the behaviour of a stiff fluid is comparable with that of the anisotropic shear - the ``matter that matters'' near to a singularity in the BKL scenario.

 To solve our equations for general $\xi$, we set the boundary condition such that the singularity is at $\eta=0$ and find
 \be u' = \frac{N}{D+N} \left\{\frac{N \coth \left[\frac{N \xi \eta }{2 \sqrt{D N (D+N-1)}}\right]}{\sqrt{D N (D+N-1)}}-1\right\} \label{eq:iso2}\ee
with energy density given by
\be \rho = \frac{N \exp[-2(u+v)]}{D (D+N) \left\{\cosh \left[\frac{N \xi \eta}{\sqrt{D N (D+N-1)}}\right]-1\right\}} \ee

The case of a cosmological constant is a subset of these solutions when $\xi=-2$. The $N+D$-dimensional deSitter solution is maximally symmetric (i.e. isotropic); for these solutions $C=0$, and we recover $u^d=v^n=\exp(\Lambda t)$ for some constant $\Lambda$. This was examined for other curvatures in \cite{Yearsley:1996yg}.


We now turn our attention to matter that has anisotropic pressures. The $N=1, \; D=2$ case was treated explicitly in \cite{Sloan:2016efd}. Taking $(w_y-1)$ times equation \ref{eq:eom1} and subtracting $(w_x-1)$ times equation \ref{eq:eom2} we find a generalization of equation \ref{eq:td} that can be expressed as:
\ba e^{-(v+u)} \f{d}{dt} \left[e^{v+u}\right. &&\left.\left(\f{[1+(D-1)w_x-D w_y]}{D} \dot{v}\right. \right. \nonumber \\
& &\left. \left. -\f{[1+(N-1)w_y -N w_x ]}{N} \dot{u}\right) \right] = 0 \nonumber \\ \ea
It is convenient to write $\xi=w_x-1$ and $\delta = w_x-w_y$. The equation of motion for $u$ thus becomes:
\ba \label{eq:aniso} u'' &+& \frac{(D+N) \xi^2 -2D\delta\xi -(N-1)D\delta^2}{2N (\xi-D\delta)} u'^2  \\
&+& (\xi-\delta) u' +\frac{N(D-1)(\xi-D\delta)}{2D(N+D-1)}=  0\ea
 We can solve this equation to find 
  \be u' = u'_o [(\xi-\delta)+ \mu (\xi-D \delta)  \coth (\f{ (\xi-D\delta) \mu\eta}{2})] \label{eq:anisosol} \ee
where the constants are
 \ba u'_o &=&\frac{N (\xi -D \delta )}{\xi ^2 (D+N)-2 D \delta  \xi -D (N-1) \delta ^2 }
 \nonumber \\ \mu&=&\frac{N}{ \sqrt{D N (D+N-1)}}.
 \ea 
and once again find the energy density:
\be \rho = u'_o\f{e^{-2(u+v)} (\xi+D\delta) }{D[1-\cosh(2 \mu \eta)]} \ee

Note that the dependence on $N$ only enters through the value of $u'_o$. Furthermore, this is positive for positive $u'_o$ when $\xi+D\delta > 0$, and this there is a strong interplay between the number of large dimensions $D$ and the matter anisotropy $\delta$. In the isotropic limit, $\delta=0$ and $\xi$ is never positive, therefore we require a strong density contrast if the initial singularity is to be at vanishing $u$. We use this to analyse the turning points of $u$ by setting $u'=0$ in equation \ref{eq:aniso}. Here we see that the sign of $u''$ is determined to be the opposite of that of $(D-1)\delta+\xi$. We will examine this relationship further in the next section, where we consider specific solutions with $D=3$ and $N \rightarrow \infty$.

Let us now consider the case of multiple internal manifolds. In such a situation if the matter content is isotropic, we can use equation \ref{Constants} to reduce the problem to that of a single ODE, the equivalent situation to equation \ref{eq:iso1}. Without loss of generality, let us label the volume and dimension of the manifold in question by $v=v_a$ and $D=D_a$, with the remaining manifolds have volumes $u_i$ and dimensions $N_i$, with $N=\sum N_i$ the codimension of the manifold in question, and $C=\sum C_{ia}$ the total of the constants relating the expansion of the manifold we have chosen and the remaining manifolds. The general form is:
\ba & 0= \frac{D+N-1}{D} v'' + \xi \left( \frac{2N+D}{2D} +\frac{N(N+1) +2 N_i N_j}{2D^2} \right) v'^2 \nonumber \\ 
  & + \xi \left(C\frac{D+ N-1}{D} +\frac{C_i N_j}{D} \right) v' + \xi \left(\sum_{k \neq a} \frac{(N_k-1)C_k^2}{2N_k} + C_i C_j \right)\label{GenIso}  \ea
herein the subscripts $i$ and $j$ should be summed over $i<j$ with neither being $a$. It is immediately apparent from this that if there is only one manifold of high dimension, this corresponds closely with equation \ref{eq:iso1}:
\be v''+\frac{\xi}{2} v'^2 + \xi C v' + \xi \left(\sum_{k \neq a} \frac{(N_k-1)C_k^2}{2N_k} + C_i C_j \right) =0 \ee
The only qualitative difference that has come about from introducing multiple manifolds in this way is that the final term can be negative. Essentially such a change can be absorbed into the choice of sign of $\eta$. 

If there are multiple high dimensional manifolds our treatment must adjust. To make life simple, we will choose to order our manifolds by their number of dimensions, and consider the case where 
\ba \frac{N_j}{D} \rightarrow \begin{cases} &\lambda_i \quad j<m \\
								&0 \quad j \geq m \end{cases} \ea
In such a case we find that the effect of the infinite dimensional manifolds upon one-another is to act as a single manifold with an extra anisotropy term that contributes only to the constant term in the ODE. Defining $\nu = (1+\sum_j \lambda_j) v$ (a rescaling of the log of the volume of the manifold of highest dimenison) we find:
\be \nu''+ \xi \frac{ \nu'^2}{2} + \xi \nu' + \xi \left(\sum_k \frac{N_k-1}{2N_k} C_k^2 + C_i C_j \right)\ee
where again the indices $i$ and $j$ should be summed for $j>i$. In such cases we can use the results above (under the substitution of $\nu$ for $v$) to describe the full set of solutions. This simple relationship can be explained physically by the fact that the product of two tori is a torus, and thus in considering multiple internal manifolds each of which is toroidal, the only difference with considering a single torus (with dimension the sum of the tori's dimensions) is that there is there can be some anisotropic expansion, and this is entirely contained in the $C_i$. 

\section{$D=3$, $N\rightarrow\infty$}
\label{sec:infinity}

We now proceed to construct our infinite dimensional cosmologies, focusing on $D=3$. If we first focus on the vacuum solutions (what is, in effect, a higher dimensional generalization of the Kasner solution) we find that 
$n_x=\pm1/\sqrt{3}$ while $n_y=0$. Is this behaviour more general, i.e. can $N\rightarrow\infty$
stabilise the extra dimensions in a more general setting?  A full derivation of the solutions in general dimension is given in appendix \ref{GenSol} with a discussion about stabilization in section \ref{sec:stabilization}. Here we will present the results in the specific case $D=3$, $N\rightarrow \infty$. One can recover exact, proper time evolutions in certain conditions for isotropic matter. As discussed in section \ref{sec:infinite} there are three matter types of interest in the infinite dimensional limit - stiff fluids, dust and cosmological constants. From our equation of motion, we see that the stiff fluid solutions are all given by $v \sim A \log(t), u \sim B \log(t)$. One branch of these has $A=0$, the other $A+B=1$. Again these are the generalizations of the locally rotationally symmetric Bianchi I solutions with stiff fluids to higher dimensions. In the former case, the three-dimensional background space is static, in the latter we can fix the internal manifold only for a scalar field whose energy density is sufficiently large (again, analogous to the Bianchi cosmologies). In such a case $B=0$ and thus the 3D space-time will appear to have a scale factor that grows as $a \sim t^\frac{1}{3}$
Let us examine first dust solutions ($\xi=-1$). Here we see
\ba \dot{u}&=& \f{12-2t}{12-t^2}\\
      \dot{v}&=& -\f{12}{12-t^2} \ea
with energy density 
\be \rho=\f{2}{t^2-12} \ee
We should note here that this solution is only really valid for $t^2>12$ as the other regions are negative energy solutions of the equations. We can integrate the equations of motion to find 
\be a = a_o \left(\f{t}{t+4\sqrt{3}}\right)^{\f{1}{\sqrt{3}}} \ee
The equation of motion for $b$ is of itself trivial in this limit, but the volume, $b^N$ is given by
\be b^N = b_o^N t^{1-\sqrt{3}} \left(t+4 \sqrt{3}\right)^{\sqrt{3}+1} \ee

The case of a cosmological constant ($\xi=-2$) is somewhat complicated by the existence of anisotropic expansion - we no longer have only the deSitter solution to consider. Solving the more general equations we find:

\ba \dot{u}&=& \f{1}{\sqrt{3}} \coth (\f{t}{\sqrt{3}}) + \text{csch}(\f{t}{\sqrt{3}})\\
      \dot{v}&=& -\text{csch}(\f{t}{\sqrt{3}}) \ea
with constant energy density, $\rho=1/6$. Note that due to the normalizations used, we find only a single solution in each instance whereas a full set of solutions should be given for general energy density. One can generate a full set of solutions from ours by rescaling $\{u,v\}\rightarrow\{\alpha u, \alpha v\}$ which scales the energy density $\rho \rightarrow \alpha^2 \rho$. On examining these solutions we find that $\dot{v}\rightarrow 0$ as $t\rightarrow \infty$. This seems to stand in contrast to what we would expect from our experience dealing with finite dimensional spacetimes, in which the cosmological constant leads to exponential expansion. The reason can be traced to insisting that $\rho$ be finite. The energy density in a deSitter spacetime relates to the expansion of the volume of the spatial slice, so as we take the large dimensional limit holding energy density fixed, the expansion of each individual scale factor tends to zero. Since $v$ represents a finite dimensional volume, and $u$ the infinite limit, it is thus unsurprising that $v$ becomes fixed. The scale factor of the three dimensions follows
\be a= a_o \tanh \left(\frac{t}{2 \sqrt{3}}\right)^{\f{1}{\sqrt{3}}} \ee
Again, the equation for $b$ is trivial and  the volume $b^N$ is given by:
\be b^N=b_o^N \sinh \left(\frac{t}{\sqrt{3}}\right) \tanh \left(\frac{t}{2 \sqrt{3}}\right)^{\sqrt{3}} \ee

It is obvious from these solutions that the late time behaviour of the three large dimensions is that they tend towards a constant size. In fact this will be the case for all isotropic matter; as $u'>0$ for all these solutions at all times, any perfect fluid with a pressure greater than that of the cosmological constant (i.e $P>-\rho$) the diffusion under expansion of the extra dimensions will cause the energy density to asymptote to zero, the vacuum solution. Furthermore, the anisotropy tends to zero at late times and so the dynamics approaches that of an infinite dimensional isotropic universe. In this case although the total volume may vary, the change in the scale factor must be infinitesimal to render this finite, a consequence of the finiteness of energy following a similar argument to that of the cosmological constant above. To see this more directly, consider the isotropic limit of \ref{eq:en1}. As the universe isotropizes, $\dot{v} \rightarrow \f{D}{N} \dot{u}$. Thus if $\rho$ is finite, then $\dot{u}$ is finite and hence $\dot{v} \rightarrow 0$. 

Let us now turn our attention to anisotropic matter. There is a particular set of solutions wherein $3w_y=2w_x+1$ which satisfy $\ddot{u}=\dot{u}=0$, and thus can lead to an exactly stable extra-dimensional space. These are discussed in section \ref{sec:stabilization}. Let us consider here a different solution, specifically where $w_x=0, w_y=-1$. This is chosen as an example of how the extra dimensions can have different late time behaviour to those of the 3 background dimensions. In such cases, $\xi=-\delta=-1$ and thus equation \ref{eq:anisosol} simplifies:
\ba u'&=& 4-\f{8}{\sqrt{3}}\coth(\f{4\eta}{\sqrt{3}}) \\
    v'&=& -2+\f{6}{\sqrt{3}}\coth(\f{4\eta}{\sqrt{3}}) \ea
Although the resulting relationship between $t$ and $\eta$ is everywhere invertible, as it is a bijection, the closed form for $\eta(t)$ does not simplify globally. However, at large $t$, we can determine the asymptotics to our solution, obtaining 
\be \eta=\f{\sqrt{3}}{1+\sqrt{3}} \log(\f{4t}{3-\sqrt{3}}) \ee
Together with the above, we see that the asymptotic behaviour is to find that $\dot{u}$ is monotonic decreasing towards zero everywhere, and $\dot{v}$ increases before tending towards zero. The volume of the extra dimensions is large but shrinks and asymptotes to a constant, whereas the 3-dimensional space expands over time. 

\section{Stabilization}
\label{sec:stabilization}
Outside of the vacuum ($\rho=0$) or stiff fluid ($w_x=w_y = 1$), it is not possible to have the dynamics of the extra dimensions stabilized by isotropic matter. In the case of a single internal manifold, this can be seen from our initial equations of motion - equations \ref{eq:eom1}. Setting $\dot{u}=\ddot{u}=0$ and $P_x=P_y$ forces $\dot{v}=\ddot{v}=0$ which contradicts the Hubble equation \ref{eq:en1} unless $\rho=0$. Now consider the case of multiple internal manifolds with (possibly) differing scale factors. A similar argument to the above shows that not all the internal manifolds can have constant volume unless $\rho=0$ or the matter is a stiff fluid. The conservation law of equation \ref{Constants} shows that $v'_i=0$ is only possible when $v'_j$ is constant for all other $j$. 

 Therefore if we wish to stabilize our extra dimensions for general matter types (e.g. for the dark energy epoch we currently observe) we must consider anisotropic matter. We can see interesting behaviour of our anisotropic system by examining equation \ref{eq:aniso} directly. When $\xi=D\delta$ this forces $u'=0$. Note that this is the same condition as is implied by the consistency of our equations of motion \ref{eq:eom1} in the case where $\dot{u}=\ddot{u}=0$. We can solve our equations of motion easily, here yielding $\dot{v}=\f{3}{C+(2+w_x)t}$. This indicates that having a large number of stable extra dimensions gives an accelerated expansion at early times ($t\rightarrow 0$) with matter pressures taking over at later times. In general, achieving inflation with extra dimensions is a difficult task \cite{Burd:1988ss}.

This leads us to consider the following scenario; suppose that our matter consists of a mixture of fluids which are represented by effective equations of state at a given time. Then both $\xi$ and $\delta$ are evolving as universe expands. Thus this mixture may encounter the situation described, and the extra dimensions will (momentarily) be frozen. At this point we find that $v'=1$, and equation \ref{eq:aniso} sets $u''=0$ also (the pole in the coefficient of $u'^2$ is first order and met with a second order zero). We thus encounter a way in which our system can stabilize. As a simplified example, we consider the limit of infinite extra dimensions added to a three dimensional background, and take the three dimensional space to be large, and thus our dynamics are dominated by fluids for which $w_x \approx -1$. If we let the effective $w_y=-1/3 + \epsilon$ we see that around the stationary solution $u' \propto \epsilon$ and hence if our effective $w_y > -1/3$ (i.e. $\epsilon > 0$ then $u'$ is positive, which should in turn reduce the effective $w_y$ until it reaches $-1/3$, whereas if $\epsilon <0$ the opposite happens. Hence our system has a stable equilibrium in which the extra dimensions are fixed in scale. Upon reaching this scale there may be a time of accelerated expansion mimicking the cosmological constant in the three large dimensions, however this will decay.

Note that our system has only required that there exists an \emph{effective} $w_y = -1/3$ - this can easily be achieved by a mixture of, for example, the cosmological constant and matter that behaves like dust in the extra dimensions and cosmological constant in the normal three. The freezing happens because of the interactions between the pressures and energy density, as can be seen directly from equations \ref{eq:eom1} without any need to make the assumption of perfect fluids in the expansion; one can simply \emph{define} the (possibly time dependent) effective equation of state be $P=w\rho$. What we have done is essentially to posit that under a perturbation about this point in phase space, there is a a negative correlation between the effective $w_y$ and $u$, since fluids with lower $w_y$ will be diluted less as $u$ expands. This is in agreement with the usual behaviour of cosmology in which fluids with lower values of $w$ dominate at large scales, and high $w$ at low scales. This negative correlation means that the equilibrium is stable. It is interesting to note that this behaviour is qualitatively independent of $N$, the number of extra dimensions - any number of extra dimensions can be stabilized by the correct type of matter. 

\section{Discussion}
\label{sec:discussion}
In this paper we have taken the first steps in studying the dynamics of an infinite dimensional universe. We have proposed a simple approach for determining the equations of motion in the Kaluza-Klein construction which allows us to study space with an arbitrary numbers of dimensions. The process laid out in section \ref{sec:eom} generalizes to arbitrary topologies, but here we have restricted ourselves to Euclidean tori.

We have looked at vacuum solutions (the analogue of the Kasner solution in higher dimensions), an isotropic perfect fluid and an anisotropic fluid and found interesting and novel behaviour. In particular we found that in the $N\rightarrow\infty$ limit, the extra dimensions are stabilized in the vacuum case. We showed that property is lost once one adds a non-zero energy momentum tensor although we discussed some ideas of how to recover it. While some familiar properties of the cosmological solutions remained in the $N-\rightarrow\infty$ limit, the overall behaviour of the large dimensions (i.e. the 3-dimensional space) is markedly different than that of a strictly 4-dimensional universe. At a first glance, from the background dynamics alone, it is unlikely that we live in an infinite dimensional universe although a more general analysis considering different geometries and sources for energy-momentum need to be considered.

This is a first analysis in what one might consider the simplest higher-dimensional universe: the Kaluza-Klein construction. But the past two decades have brought to the fore an altogether different approach to higher-dimensions involving 3 dimensional structures, i.e. branes. Some of the machinery for p-branes in spaces of arbitrary dimensions has already has already been established \cite{Charmousis:2005ey} but an analysis such as the one done here remains to be done.

It would also be interesting to link the results of \cite{Emparan:2013moa} with those found in this paper. Specifically, it would interesting to see how the ultra-localization that the authors found there plays itself out in, for example, the cosmological solutions for dust-like matter. One approach is to attempt a bottom up construction of cosmology, as advocated in \cite{Clifton:2013jpa} that could link the smaller scales with global evolution.

Finally, and pragmatically, one might consider the $N\rightarrow\infty$ limit as a useful calculational tool. By taking expansions in $1/N$ it may be possible to find approximate, yet accurate solutions for Kaluza-Klein theories with a moderate number of extra dimensions - for example string theory or super gravity invoke $N=6$ or $N=7$ for which corrections may be sufficiently small. Regardless, the analysis of sections \ref{sec:eom} and \ref{sec:solutions} is exact for any $N$ and $D$.

\textit{Acknowledgements ---} PGF was inspired to look at $N\rightarrow\infty$ by comments in \cite{Deser:1998ed}. We acknowledge discussions with J. Noller and J. Scargill. PGF acknowledges support from Leverhulme, STFC, BIPAC and the ERC. DS acknowledges support from the John Templeton Foundation. The authors are indebted to John Barrow, Stanley Deser, Alex Feinstein and Daniel Farid Litim for their comments.

\appendix
\section{General Isotropic Matter Solutions: Dust and Cosmological Constant }
\label{GenSol}
To obtain a general solution, we note that it is simpler to stay with the first time derivatives of variables throughout our transformations between $t$ and $\eta$: 
From equations \ref{eq:const} and \ref{eq:iso2} we find that $u'+v'$ takes a simpler form, and so we can find the relationship between $\eta$ and $t$ simply:
\be \label{eq:Dedt} \sinh \left[\frac{N \xi \eta }{2 \sqrt{D N (D+N-1)}}\right]^{2/\xi} d\eta =dt \ee
from which we can generally recover $\eta$ as a function of $t$. Unfortunately, for general $\eta$ this involves inverting a confluent hypergeometric function which supports multiple branch cuts leading to different solutions in different branches. However, in the case where $\xi=-1$ (i.e. dust) we find that this results in 

\be t= -\frac{2}{\mu} \coth \left( \f{\mu \eta}{2} \right)\ee

which is easily inverted for t. We set $\mu = \f{N}{\sqrt{ND(N+D-1)}}$ for brevity. Using this together with $\dot{v}=v' \dot{\eta}$ and equation \ref{eq:Dedt} we see

\ba \dot{u}&=&\frac{4 D N (D+N-1)-2 N^2 t}{(D+N) \left(4 D (D+N-1)-N t^2\right)} \\
      \dot{v}&=& -\frac{2 D N (2 (D+N-1)+t)}{(D+N) \left(4 D (D+N-1)-N t^2\right)}\ea 

These can be integrated to give:

\ba u&=& \frac{\mu D \log \left(4 D (D+N-1)-N t^2\right)-2 N \tanh ^{-1}\left(\frac{\mu t}{2}\right)}{\mu(D+N)} \\
      v&=& \frac{\mu N \log \left(4 D (D+N-1)-N t^2\right)+2 N \tanh ^{-1}\left(\frac{\mu t}{2}\right)}{\mu(D+N)} \ea

It is clear that these can be split into the growth of the isotropic volume  and the anisotropy easily. The measure of anisotropic expansion is 

\be \Sigma = \f{\dot{u}}{N}- \f{\dot{v}}{D} = \frac{4 (D+N-1)}{4 D (D+N-1)-N t^2} \ee

This is singular at the origin, but reduces as the matter presses the system into isotropy. 

We can perform a similar analysis by considering the case of the cosmological constant ($\xi=-2$). Here we follow the above to obtain:

\ba \dot{u}&=&  \frac{N}{D+N} \frac{1}{\sinh (\mu t)} \left( \mu \cosh(\mu t) +1  \right) \\
      \dot{v}&=&  \frac{N}{D+N} \frac{1}{\sinh (\mu t)} \left( \f{D}{N} \mu \cosh(\mu t) -1  \right)\ea 
Which has a constant energy $\rho = \f{N}{2D(N+D)}$. 
This solution asymptotes to the isotropic limit. The measure of anisotropy in expansion:

\be \Sigma = \f{\dot{u}}{N}- \f{\dot{v}}{D} = \frac{1}{d \sinh(\mu t)} \ee

At early times this is very large, diverging at the origin, but quickly tends to zero as time goes on, and we return to an isotropic exponential expansion as the system asymptotes to deSitter expansion. 

\section{Generalized Billiards}
\label{billiards}
The cosmological billiards approach rewrites the metric in a manner such that the anisotropies have the form of scalar fields evolving on a shape potential which arises from the Ricci scalar. We shall here present a small generalization of this approach to include the block-diagonal metric we use. We identify sets of the translational one-forms under which our spatial slices are homogeneous to form isotropic subspaces, and hence our approach is equivalent to taking the cosmological billiards approach on a $M$-dimensional spatial manifold and enforcing equalities between some of the scalar fields. We express our metric in the form;
\be ds^2 = -dt^2 + e^{2 \v} \sum_i e^{2\s_i} ds_i^2 \ee
wherein the final shape parameter $\s_q = - \sum_{i<q} \s_i$. As before, the $ds_i$ are $D_i$-dimensional homogeneous isotropic manifolds of unit volume. However, in this form we have explicitly captured the behaviour of the entire volume of space in $\v$. The Ricci scalar will split into the product of a term purely in $\v$ and a term $R_s[\vec{\s}]$ which depends only on the shape parameters (called the `shape potential'). Our Langrangian becomes
\ba \L_g = e^{\v} ( &-&\left[\sum_i \frac{D_i-1}{D_i} + q(q-1)\right] \dot{\v}^2 \nonumber  \\
                                   &+& \sum_{i<q} \left[\frac{D_i-1}{D_i}  + \frac{D_q-1}{D_q} -2\right] \dot{\s_i^2} \nonumber \\
                                   &+& 2 \left[\frac{D_q-1}{D_q}-1\right] \sum_{i \neq j} \dot{\s_i}\dot{\s_j} \nonumber \\
                                   &+& 2\sum_{i<q}\left[ \frac{D_i-1}{D_i} -\frac{D_q-1}{D_q}\right]\dot{\v}\dot{\s_i} \nonumber \\
                                   &+& e^{\frac{2-M}{M} \v} R_s[\vec{\s}]) \ea
In the case where we do not impose each of the isotropies on the spatial manifolds the above reduces to the usual cosmological billiards Lagrangian, setting each $D_i=1$ and $q=M$. This setup is a starting point for investigation of the shape dynamics of such a system, which is beyond the scope of the current work. As we are here primarily interested in the case of one large spatial manifold and many small ones, it is more convenient to use the parametrizations given above, rather than dealing with a mixture of $\v$ and $\s_i$ terms. Nevertheless, we can gain insight into our system by considering the behaviour of this system. In particular, we note that there exist geometries for which $R_s$ is independent of one or more of the shape parameters. A trivial example of this is that of the product of tori, but this can happen in the case of more complicated systems. When this is the case, the Euler-Lagrange equations immediately reveal a conserved quantity, $\frac{\delta \L}{\delta_{\dot{\s_i}}}$. In the toroidal case there correspond exactly to the $C_{ij}$ of equation \ref{Constants}.

\section{General Einstein Equations}

In section \ref{sec:eom} we presented the Einstein equations that arise when the spatial manifold is the product of two tori. Here we will give the equations in full generality. These can be easily derived from the Lagrangian in equation \ref{UVLag} by singling out one particular spatial manifold at a time to consider as the `background'. In doing so we find the equations arising from the spatial eigenvalues of the Einstein tensor (the pressures) are given;
\ba &-&P_1 = \frac{D_1 -1}{D_1} \ddot{v_1} + \frac{D_1-1}{2D_i} \dot{v_1}^2 + \sum_{1<j<k} \dot{v_j} \dot{v_k} +\frac{D_1-2}{2D_1} e^{-\frac{2v_1}{D_1}} {}^1 R \nonumber \\ 
              &+& \sum_{j>1} \left[ \frac{D_1-1}{D_1} \dot{v_1}  \dot{v_j} +\frac{D_j+1}{2D_j} \dot{v_j}^2  +\ddot{v_j} + \frac{e^{-\frac{2v_j}{D_j}} {}^j R}{2} \right] \ea
with $P_2$ etc given cyclically. Again we find the Friedmann equation from the Hamiltonian of the system;
\be \rho = \sum_i \left(\frac{D_i-1}{2D_i} \dot{v_i}^2 + e^{-\frac{2v_j}{D_j}} {}^j R \right)+ \sum_{i < j} \dot{v_i} \dot{v_j} \ee    
Here we reiterate the observation that the manner in which we approach the limit of taking a large number of dimensions in our spatial manifold can lead to different dominating behaviours - if we take the number of manifolds $q$ to be large, the cross between $\dot{v_i}\dot{v_j}$ grows as $q^2$ whereas the others grow linearly in $q$, and hence are insignificant. However, if we take the limit of a large number of dimensions $n$ in a homogeneous, isotropic spatial manifold, we expect that the curvature term ${}^i R$ to grow as $n^2$ and thus dominate. 
            

\bibliographystyle{apsrev4-1}
\bibliography{Infinity}

\end{document}